\documentclass[aps,prl,amsmath,amssymb,showkeys,showpacs,preprintnumbers,nofootinbib,twocolumn]{revtex4}

\usepackage{mathrsfs}
\usepackage{graphicx}
\usepackage[colorlinks=true]{hyperref}
\usepackage{color}

\newcommand{\bw}{\begin{widetext}}
\newcommand{\ew}{\end{widetext}}
\newcommand{\be}{\begin{equation}}
\newcommand{\en}{\end{equation}}
\newcommand{\bea}{\begin{eqnarray}}
\newcommand{\ena}{\end{eqnarray}}

\def\vec{\mathbf}

\def\etal{{\it et.al.}}

\def\d{{\rm d}}

\begin{document}

\title{Hybird of Quantum Phases for Induced Dipole Moments}

\author{Kai Ma}
\email{makainca@yeah.net}
\affiliation{School of Physics Science, Shaanxi University of Technology, Hanzhong 723000, Shaanxi, P. R. China}

\begin{abstract}
The quantum phase effects for induced electric and magnetic dipole moments are investigated. It is shown that the phase shift received by induced electric dipole has the same form with the one induced by magnetic dipole moment, therefore the total phase is a hybrid of these two types of phase. This feature indicates that in order to have a decisive measurement on either one of these two phases, it is necessary to measure the velocity dependence of the observed phase. 
\end{abstract}



\date{\today}

\maketitle

The interference based on the wave nature of matter has been the most important test of quantum mechanics. The high sensitivity of matter-wave interference experiments to external perturbations has been viewed as the most powerful tequniqe in accurately measuring internal properties of composite particles. Particularly, the topological phase effects on matter wave in interference experiment were the most unique. The first topological phase shift was predicted by Aharonov and Bohm (AB) in 1959 \cite{Aharonov-Bohm}, and an excellent agreement was found between the measured phase shift and the theoretical prediction. This effect has also been tested at the macroscopic level \cite{TimeDelay}. Nearly same effect was proposed for neutral spinor particle with non-vannishing magnetic dipole moment by Aharonov and Casher (AC) in 1984 \cite{Aharonov-Casher}. This effect has also been confirmed by the neutron and atom interference experiments \cite{Cimmino:neutron,Allman:sab, Sangster:atom}. Possible topological phase due to the interaction between electric dipole moment of a neutral spin-1/2 particle and magnetic fields generated by magnetic monopole was suggested by He and McKellar \cite{He-McKellar} and a year later independently by Wilkens \cite{Wilkens}. This effect is related to the AC effect through the possible symmetry---dual symmetry in electromagnetic theory. However, this effect has not been observed due to the absence of magnetic monopole. 

On the other hand, it was pointed out that when both the electric and magnetic fields are applied, possible phase shift can be accumulated by atom with non-vanishing electric polarizability \cite{Wei:1995}, and explored by J. P. Dowling {\etal}  in further \cite{Dowling}. However, there are some differences in the phase shift effects for permanent dipole and induced dipole moments. One of these is the influence of the inevitable dynamical property of induced dipole moment, which was pointed out in first by Wilkens \cite{Wilkens:time-dipole} and further discussed by Horsley and Babiker \cite{Horsley:2007}. It was pointed out that the time varying of the dipole can result in a nontrivial phase shift. Recently, S. Lepoutre \etal \cite{Lepoutre:exp-hmw} reported that they have detected a phase shift effect for an electrically polarized atom by using an atom interferometer. In their experiment, an electric dipole moment $\vec{d}=4\pi\varepsilon_{0}\alpha\vec{E}$ is induced by an applied electric field $\vec{E}$ on $^{7}\text{Li}$ atom \cite{Miffre:elecdipo}. Then, its interaction with applied magnetic field $\vec{B}$ induces an nontrivial phase shift,
\begin{equation}\label{lepoute-phase}
     \Delta\phi_{\rm E}
    =\frac{1}{\hbar}\oint\d\vec{r}\cdot(\vec{d}\times\vec{B}).
\end{equation}
However, the measured value of the phase differs by $31\%$ from its theoretical value. While this difference possiblly comes form some as yet uncontrolled systematic errors, we will show that it may be result of nonzero magnetic polarizability. This possibility originates in the hybrid of quantum phases for electric and magnetic dipole moments. For ``fundamental" particles, such as electron and neutron, the electric dipole moment is extremely small comparing to its magnetic dipole moment, then the phase shift is dominated by the magnetic one. Nevertheless, the mixing may be significant for atom with large electric dipole moment. In the following discussions, we will focus on the phase effects for atom.

Let us first consider a neutral particle of mass $m$ with nonzero magnetic polarizability $\kappa$. When it moves in electromagnetic field with velocity $\vec v$ in the library frame, it will be polarized and hence carry an induced magnetic dipole moment. In the rest frame of particle, it is given by,
\begin{equation}\label{induced-magnetic-dipole}
    \vec{m}
    =\kappa(\vec{B}-\frac{\vec{v}}{c}\times\vec{E})\,.
\end{equation}
This dipole moment constitutes two parts. The first one $\vec{m}_{\mathcal R}\equiv\kappa\vec{B}$ is due to the standard magnetic polarization. The second one $\vec{m}_{\mathcal M}\equiv-\kappa{\vec{v}\over{c}}\times\vec{E}$ stands for the polarization due to the effective magnetic fields felted by the particle in its rest frame. This induced magnetic dipole couples to the external magnetic and electric fields, and the Lagrangian can be written as,
\begin{equation}\label{classical-lagrangian-induced-md}
    \mathcal{L}
    =\frac{1}{2}m\vec{v}^2
     +\frac{\kappa}{2}(\vec{B}-\frac{\vec{v}}{c}\times\vec{E})\cdot(\vec{B}-\frac{\vec{v}}{c}\times\vec{E}).
\end{equation}
From this Lagrangian one can obtain the canonical momentum,
\begin{equation}\label{canonical-momentum}
     \vec{P}
    =\frac{\partial\mathcal{L}}{\partial\vec{v}}
    =m\vec{v}+\frac{\vec{m}_{\mathcal R}}{c}\times\vec{E}+\frac{\vec{m}_{\mathcal M}}{c}\times\vec{E}.
\end{equation}
The last two terms of this canonical momentum give rise to the minimal coupling which can be seen in the corresponding Hamiltonian,
\begin{equation}\label{hamiltonian}
    \mathcal{H}
    =\frac{1}{2m}(\vec{P}-\frac{\kappa}{c}\vec{\mathcal A})^2+\kappa\mathcal{V}~,
\end{equation}
where
\begin{equation}\label{effective-vector-potential}
     \vec{\mathcal{A}}
    =(\vec{B}-\frac{\vec{v}}{c}\times\vec{E})\times\vec{E}~,~~
    \mathcal V=-\frac{1}{2}\kappa\vec{B}^2
\end{equation}
Effectively, this Hamiltonian describes the interaction of a ``matter particle" with effective charge $\kappa$ and ``electromagnetic field" with vector potential $\vec{\mathcal{A}}$ and scalar potential $\mathcal V=-\vec{B}^2/2$. Just as in the AB and AC effect \cite{Ma:Uone}, possible nontrivial phase shift can be generated by requiring that in the region of particle moving the effective magnetic and electric fields vanishing, 
\begin{equation}\label{condition1}
    \vec{\mathcal{B}}_{m}
    =\vec{\nabla}\times\vec{\mathcal{A}}
    =0~,~~
    \vec{\mathcal{E}}_{m}
    =-\vec{\nabla}\mathcal{V}
    =0~.
\end{equation}
Then, the nontrivial phase shifts are,
\begin{equation}\label{phase-magnetic-induced1}
      \Delta\phi_{\rm S}
    =\frac{\kappa}{\hbar c}\oint\d\vec{r}\cdot\vec{\mathcal A}~,~~
      \Delta\phi_{\rm T}
    =-\frac{\kappa}{\hbar}\oint\d t{\mathcal V}~.
\end{equation}

If the electric field is constant over the two arms of interferometer, the second term in (\ref{phase-magnetic-induced1}) will be zero due to cancellation between the contributions of two arms. The first term which is proportional to the magnetic and electric fields can generate a phase effect analogy to the ordinary AC effect with the replacement $\vec{\mu}_{\rm r}=\kappa\vec{B}$. It is worthy to point out that, at a first glance, this partial phase shift is not dispersionless due to the explicit dependence of particle velocity. In fact, however, it is dispersionless. This feature can be understand from that while one shifts the velocity of particle, the contribution from this change are attributed to the first term in (\ref{phase-magnetic-induced1}).

By requiring the external electromagnetic field satisfies the condition (\ref{condition1}), this effect can be force free. The simplest solution is that both electric and magnetic field are constant. However, as we have mentioned, to generate a nontrivial and measurable phase shift for enclosed path, the effective magnetic field $\vec{\mathcal{B}}_{m}$ must be nonzero, $\vec{\mathcal{B}}_{m}\neq0$, at some space region enclosed by the particle trajectory. With straightforward calculation, under the condition $\vec{\nabla}\cdot\vec{B}=0$, one can conclude that the only possible configuration is that particle trajectory encloses nonzero electric charge distributions. 

Similar results can be obtained for the electric dipole moment induced by external electric field, which is described by the Lagrangian,
\begin{equation}
    \tilde{\mathcal{L}}
    =\frac{1}{2}m\vec{v}^2 
      +\frac{\chi}{2}(\vec{E}+\frac{\vec{v}}{c}\times\vec{B})\cdot(\vec{E}+\frac{\vec{v}}{c}\times\vec{B})~.
\end{equation}
Then, the nontrivial phase shifts are,
\begin{equation}\label{phase-electric-induced1}
      \Delta\tilde\phi_{\rm S}
    =\frac{\chi}{\hbar c}\oint\d\vec{r}\cdot\tilde{\vec{\mathcal A}}~,~~
      \Delta\tilde\phi_{\rm T}
    =-\frac{\chi}{\hbar}\oint\d t\tilde{\mathcal V}~.
\end{equation}
where
\begin{equation}\label{effective-vector-potential}
     \tilde{\vec{\mathcal{A}}}
    =-(\vec{E}+\frac{\vec{v}}{c}\times\vec{B})\times\vec{B}~,~~
    \tilde{\mathcal V}=-\frac{1}{2}\vec{E}^2
\end{equation}
Effectively, this Hamiltonian describes the interaction of a ``matter particle" with effective charge $\chi$ and ``electromagnetic field" with vector potential $\vec{\tilde{\mathcal{A}}}$ and scalar potential $\tilde{\mathcal{V}}=-\vec{E}^2/2$. Just as in the AB and AC effect, possible nontrivial phase shift can be generated by requiring that in the region of particle moving the effective magnetic and electric fields vanishing, 
\begin{equation}\label{condition}
    \tilde{\vec{\mathcal{B}}}_{e}
    =\vec{\nabla}\times\tilde{\vec{\mathcal{A}}}
    =0~,~~
    \tilde{\vec{\mathcal{E}}}_{e}
    =-\vec{\nabla}\tilde{\mathcal{V}}
    =0~.
\end{equation}
Similar to the phase (\ref{phase-magnetic-induced1}), the second term in (\ref{phase-electric-induced1}) has no contribution for an enclosed path with constant magnetic field over the path.

For a particle which processes both nonzero the magnetic and electric polarizibilities, the combined phase shift is,
\begin{equation}\label{phase-electric-induced}
    \phi_{\rm t}
    =\phi_{\rm m} + \phi_{\rm e}
    =\frac{\kappa+\chi}{\hbar}\oint\d\vec{r}\cdot\bigg(\vec{B}\times\vec{E}\bigg)
\end{equation}

In the general case, if there are both electric and magnetic field in the space region of particle moving, the total phase for permanent electric and magnetic dipole moments is,
\begin{equation}\label{total-phase-permanent1}
    \Delta\phi_{\rm perm}
    =\Delta\phi_{\rm AC}^{\vec E}+\Delta\phi_{\rm AC}^{\vec B}+\Delta\phi_{\rm HMW}^{\vec E}+\Delta\phi_{\rm HMW}^{\vec B}\,.
\end{equation}
In the case that the particle has both nonzero the magnetic and electric polarizibilities, and one provides that the atom has induced electric dipole $\vec{d}_{p}=\chi\vec{E}$ for electric field $\vec{E}$ and induced magnetic dipole $\vec{\mu}_{p}=\kappa\vec{B}$, then the total phase can be obtained by replacing the permanent electric and magnetic dipole in (\ref{total-phase-permanent1}) with the total quantities,
\begin{equation}\label{total-dipole}
    \vec{d}=\vec{d}_{0}+\vec{d}_{p},~~
    \vec{\mu}=\vec{\mu}_{0}+\vec{\mu}_{p},
\end{equation}
where $\vec{d}_{0}$ and $\vec{\mu}_{0}$ are the permanent electric and magnetic dipole moments, respectively. Then one can get the total phase received by the atom,
\begin{equation}\label{total-phase-permanent}
    \Delta\phi_{\rm ind}
    =\Delta\phi_{\rm perm}+\Delta\phi_{mix}+\Delta\phi_{S}+\Delta\phi_{Z},
\end{equation}
where,
\begin{equation}\label{phase-mix}
    \Delta\phi_{mix}
    =\frac{\chi-\kappa}{\hbar}\oint(\vec{E}\times\vec{B})\cdot\d\vec{r},
\end{equation}
\begin{equation}\label{phase-stark-zemman}
    \Delta\phi_{S}
    =\frac{\chi}{\hbar}\oint\vec{E}^2\d t,~~
    \Delta\phi_{Z}
    =\frac{\kappa}{\hbar}\oint\vec{B}^2\d t.
\end{equation}
Usually, the Stark and Zeeman phases shifts can be adjusted to zero by arranging the experiments such that the integrands on the two path are equal. The interesting thing is the phase shift (\ref{phase-mix}).

Recently, by using an atom interferometer, S. Lepoutre \etal \cite{Lepoutre:exp-hmw} reported that they detected a nontrivial phase shift for atom $^{7}{\rm Li}$. 
The result was parametrized by eq. (\ref{lepoute-phase}) which can also be written as,
\begin{eqnarray}\label{lepoute-phase-re}
    \Delta\phi_{\rm exp}
    \nonumber
    &=&\frac{4\pi\alpha}{\hbar}\oint\d\vec{r}\cdot(\vec{E}\times\vec{B}) \\ 
    &=&\frac{1}{\hbar}\int_{\Omega}\d\vec{S}\cdot(4\pi\varepsilon_{0}\alpha\vec{B})(\vec{\nabla}\cdot\vec{E}).
\end{eqnarray}
One can clearly see that nonzero phase shift implies nonzero electric charge distribution enclosed by the trajectory of particle, and vice verse. This is just the distinctive feature of AC effect. Further, one can make an identification,
\begin{equation}\label{magnetic-dipole-identification}
      \vec{\mu}
      =4\pi\mu_{0}\alpha\vec{B}.
\end{equation}
Then this phase is exactly the phase shift for an induced magnetic dipole moment $\vec{\mu}$ with polarizibility $\kappa=4\pi\mu_{0}\alpha$. So, to identify the observed phase shift $\Delta\phi_{\rm exp}$ as the one induced by electric dipole moment, one must exclude the possibility of a nonzero induced magnetic dipole moment. In Lepoutre \etal's paper \cite{Lepoutre:exp-hmw}, the electric dipole was detected by using the atom interferometer with the following parametrization of potential enegy \cite{Miffre:elecdipo},
\begin{equation}\label{merric-d}
     U
    =-2\pi\varepsilon_{0}\alpha(\vec{E})^2,
\end{equation}
for an electric dipole moment $\vec{d}=4\pi\varepsilon_{0}\alpha\vec{E}$. And the phase shift was parametrized as,
\begin{equation}\label{merric-phase}
     \Delta\phi'
    =\frac{2\pi\varepsilon_{0}\alpha}{\hbar}\int_{\mathcal{C}}\d r\frac{(\vec{E})^2}{v}.
\end{equation}

However, assuming the atom has nonzero magnetic polarizability $\kappa$, then according to eq.(\ref{phase-magnetic-induced1}), there will be another phase shift,
\begin{equation}\label{m-electric-field}
    \Delta\phi_{\rm m}
    =-\frac{\kappa}{\hbar}\oint\d\vec{r}\cdot((\vec{v}\times\vec{E})\times\vec{E})
\end{equation}
Making a transformation
\begin{equation}\label{relation-magelec}
    \vec{d}'=\vec{\mu}'\times\vec{v}
\end{equation}
on eq.(\ref{merric-phase}), then,
\begin{equation}\label{merric-phase1}
     \Delta\phi'
    =\frac{1}{\hbar}\int_{\mathcal{C}}\d t(\vec{\mu}'\times\vec{v})\cdot\vec{E}
    =\frac{1}{\hbar}\int_{\mathcal{C}}\d t\vec{\mu}'\cdot\vec{B}',
\end{equation}
where $\vec{B}'=\vec{v}\times\vec{E}$ is the magnetic field in the rest frame of particle. This means that the observed phase shift can also be identified as the non-vanishing magnetic dipole moment $\vec{\mu}'$ with the relation $\vec{d}'=\vec{\mu}'\times\vec{v}$. Now, back to the previous identification $\vec{\mu}=4\pi\varepsilon_{0}\alpha\vec{B}$ and substitute it with $\vec{\mu}'=\vec{\mu}/2$ \cite{explan} into this relation, one has,
\begin{equation}\label{dipole}
     \vec{d}'
    =2\pi\varepsilon_{0}\alpha\vec{B}\times\vec{v}
    =2\pi\varepsilon_{0}\alpha\vec{E}',
\end{equation}
where $\vec{E}'=\vec{B}\times\vec{v}$ is the electric field in the rest frame of particle. This mean that the relation (\ref{relation-magelec}) is consistent under the identification $\vec{\mu}=4\pi\varepsilon_{0}\alpha\vec{B}$. Thus the measured quantum phase could be a mixture of the phases induced by electric and magnetic dipole moments.

In summary, we have demonstrated clearly that the total quantum phase in a general case is a mixture of the phases induced by electric and magnetic dipole moments. And this could be a possible reason of the $31\%$ offset observed in Lepoutre \etal's experiment \cite{Lepoutre:exp-hmw}.

\begin{acknowledgments}
K. M. is supported by the China Scholarship Council and the Hanjiang Scholar Project of Shaanxi University of Technology.
\end{acknowledgments}


\begin{thebibliography}{99}
\bibitem{explan}
This is because that different normalizations in papers \cite{Lepoutre:exp-hmw} and \cite{Miffre:elecdipo} have been used. This difference also can be seen in Eqs. (\ref{dipole}) and (\ref{merric-d}).
\end{thebibliography}
\end{document}